# Effective density of states profiles of heterogeneous microcrystalline silicon

Sanjay K. Ram[*] and Satyendra Kumar[!]
*Department of Physics, Indian Institute of Technology Kanpur, Kanpur-208016, India*

The steady state photoconductivity as a function of temperature and light intensity was measured on plasma deposited highly crystalline undoped hydrogenated microcrystalline silicon films possessing different thicknesses and microstructures. Different phototransport behaviors were observed experimentally in films having dissimilar microstructural attributes. This has been explained by numerical modeling to link these behaviors to different features of the proposed density of states maps of the material.

PACS numbers: 73.50.Pz, 73.50.–h, 73.61.–r, 71.23.–k, 73.50.Gr, 72.20.Jv, 81.40.Gh

Hydrogenated microcrystalline silicon (µc-Si:H) has become an unrivalled contender in the field of large area electronic device applications.[1,2] In spite of the attractive optoelectronic properties of µc-Si:H, the heterogeneous nature of its microstructure hinders a comprehensive interpretation of those properties. The recombination mechanisms and the nature of density of gap states (DOS) of µc-Si:H system are still inadequately understood, as there cannot be any unique *effective* DOS profile that would satisfy the whole microstructural range of µc-Si:H materials and explain all the intricacies involved in the transport mechanisms.[3,4] An endeavor to construct DOS profiles applicable to the µc-Si:H system would entail a study of a wide range of samples deposited under a wide array of conditions, segregable into categories on the basis of some unique microstructural and phototransport properties.[5,6,7] The DOS profiles would not only add to our knowledge of the physics of this material, but also are essential for further improvement in µc-Si:H based device technology.

Steady state photoconductivity (SSPC) is an efficient and easy technique to examine the gap states over a wider range in the bandgap, and is sensitive to both density and nature of all the defect states acting as recombination centers between the quasi Fermi levels in the band gap. Numerical modeling has been extensively used to elicit information about the recombination kinetics and explain the experimental photoconductivity results in a-Si:H,[8,9,10,11,12,13] but has been less employed in µc-Si:H.[3,5,7,14,15,16,17,18] In this letter, we report on our study of phototransport properties of a range of microstructurally different highly crystallized undoped µc-Si:H films that were classified into three types on the basis of specific microstructural attributes[19] and dark electrical transport behavior[20]. We have employed both experimental method and numerical modeling of SSPC, and have proposed the effective DOS maps of these materials.

The undoped µc-Si:H films were deposited at low substrate temperature ($T_s \leq 200$ °C) in a parallel-plate glow discharge plasma enhanced chemical vapor deposition system operating at a standard rf frequency of 13.56 MHz, using high purity $SiF_4$, Ar and $H_2$ as feed gases. Different microstructural series of samples were created by systematically varying gas flow ratios ($R = SiF_4 / H_2$) or $T_s$. We employed Raman scattering (RS), spectroscopic ellipsometry (SE), X-ray diffraction, and atomic force microscopy for structural investigations. High crystallinity of all the samples was confirmed by RS and SE measurements. SE data shows a crystalline volume fraction >90% from the initial stages of growth, with the rest being density deficit having no amorphous phase, and a reduced incubation layer thickness. There is significant variation in the percentage fraction of constituent small and large crystallite grains (SG~6-7nm and LG~70-80nm) with film growth.[19,21,22] These well-characterized annealed samples were studied for the electron transport behavior in coplanar geometry using dark conductivity ($\sigma_d(T)$, 15-450K) and photoconductivity ($\sigma_{ph}(T)$, as function of temperature, wavelength and intensity of probing light).[5,6,7] All the measurements were conducted for different film thicknesses and $R$ values. The effect of light intensity variation on SSPC was probed using above-bandgap light (He-Ne laser, λ=632.8nm) in the temperature range of 20K–324K. Photon flux $\phi$ was varied from ~$10^{11}$-$10^{17}$ photons/cm$^2$-sec using neutral density filters giving rise to generation rates of ≈ $G_L$=$10^{15}$–$10^{21}$cm$^{-3}$s$^{-1}$. In general, photoconductivity exhibits a non-integer power law dependence on carrier generation rate $G_L$ over several orders of magnitude given by $\sigma_{ph} \propto G_L^\gamma$.

The photoconductivity light intensity exponent $\gamma$ gives information on the recombination mechanism in a semiconducting material. According to Rose, $\gamma = T_c / (T+T_c)$ where $T_c$ is the characteristic energy of conduction band tail (CBT).[23,24]

The findings of the structural investigations into the microstructure and growth type of the µc-Si:H films at various stages of growth (film thickness ~50nm to ~1200nm), correlative with the dark electrical transport properties led us to segregate some common and distinct features present in the varieties of films, which we then classified into three types: *A*, *B*, and *C*. Our structural studies have shown that the percentage volume fraction of the LG ($F_{CL}$%) is a physically rational parameter that indicates the microstructural and morphological condition of our µc-Si:H films and correlates well with the conductivity behavior at both high and low temperatures.[19,20] The classification of the materials is being briefly outlined here, as it will facilitate understanding the phototransport results. The *type-A* films have small grains, high density of inter-grain boundary regions containing disordered phase, and low amount of conglomeration. Here $F_{CL}\% <30\%$, and constant activation energy ($E_a \sim 0.5$ eV) and dark conductivity prefactor ($\sigma_0 \approx 10^3$ $(\Omega cm)^{-1}$) are seen. The *type-B* films contain a fixed ratio of mixed grains in the bulk. With film growth, conglomeration of grains results in a marked morphological variation, and a moderate amount of disordered phase in the conglomerate boundaries limits the electrical transport. Here $F_{CL}\%$ varies from 30% – 45%, there is a sharp drop in $\sigma_0$ (from $\sim 10^3$ to 0.1 $(\Omega cm)^{-1}$) and $E_a$ (from $\sim 0.55$ to 0.2 eV). The *type-C* films are fully crystallized with a high value of $F_{CL}\%$ (>50%), crystallite conglomerates are densely packed, and crystallites attain preferential orientation. Here $\sigma_0$ shows a rising trend (from 0.05 to 1 $(\Omega cm)^{-1}$) and the fall in $E_a$ is slowed down (from 0.2 to 0.10 eV). Our dark electrical transport studies on these three types of $\mu$c-Si:H materials have indicated a band tail transport. Recent experimental evidence also suggests a band tail transport in $\mu$c-Si:H material similar to our material.[25]

[*]skram@iitk.ac.in (Corresponding author: S.K. Ram).
[!]satyen@iitk.ac.in

In this letter we have reported the results of phototransport studies of one representative sample from each type, since all the samples of a certain type showed similar trends. The experimental results of $\sigma_{ph}(\phi,T)$ and $\sigma_d(T)$ of all the three samples are shown in Fig. 1(a) and the temperature dependence of $\gamma$ (obtained from light intensity dependence of $\sigma_{ph}(\phi)$ at different temperatures) of the same samples is shown in Fig. 1(b). The zoomed view of significant areas of each graph is shown in their respective insets. The phototransport properties of the *type-A* material (sample #B22) shows $0.5<\gamma<1$, and $\sigma_{ph}(\phi,T)$ is essentially an increasing function of temperature. But at higher temperatures, $\sigma_{ph}$ decreases with increasing temperature, exhibiting thermal quenching (*TQ*). The solid line in this graph (Fig. 1(a)) represents the activated $\sigma_d$ behavior with $E_a=0.55$ eV over a large temperature range (~170–450K). In *type-B* material, $0.5<\gamma<1$, with $\sigma_{ph}$ increasing monotonically with temperature without any *TQ* effect as shown for sample #B23 in Fig. 1(a), where $\sigma_d(T)$ is shown by dashed line, with $E_a=0.34$ eV. In *type-C* material, $0.15<\gamma<1$ and we observe an anomalously low value of $\gamma\sim0.13$ at ~225K. The $\sigma_{ph}(T)$ monotonically increases with the increase in temperature excluding the lowest temperature region (<30K) where $\sigma_{ph}$ is nearly independent of *T*, and exhibits *TQ* effect at higher temperatures. These findings for the sample #F06 are depicted in Fig. 1(a), where $\sigma_d(T)$ is shown by dotted line, with $E_a=0.12$ eV. The qualitative analysis of these observed phototransport behaviors have been discussed in our earlier reported work,[5,6,7] and is being mentioned here in brief. In *type-A* and *type-B* material, the information about the distribution of gap states near the CBT is drawn from the exponent $\gamma$ by applying Rose model. The average width of CBT ($kT_c$) is deduced to be ~30meV and ~25-28meV in *types-A* and *B* respectively. The qualitative understanding of the phototransport properties of *type-A* material is possible in a framework of models applicable to *a*-Si:H, though each model can be applied to only a limited aspect of the $\mu$c-Si:H properties. For $\mu$c-Si:H material exhibiting behavior similar to *type-B*, symmetric band tails and a Gaussian type valence band tail (VBT) have been proposed to be responsible.[3] In *type-C* material, a very narrow width of CBT ($kT_c \sim 0.01$ eV, which is less than $kT$) exists, and therefore Rose model does not work.[13] Dangling bonds (DBs) are unlikely to cause TQ, and possibly the DOS features are similar to those of poly-Si.[26]

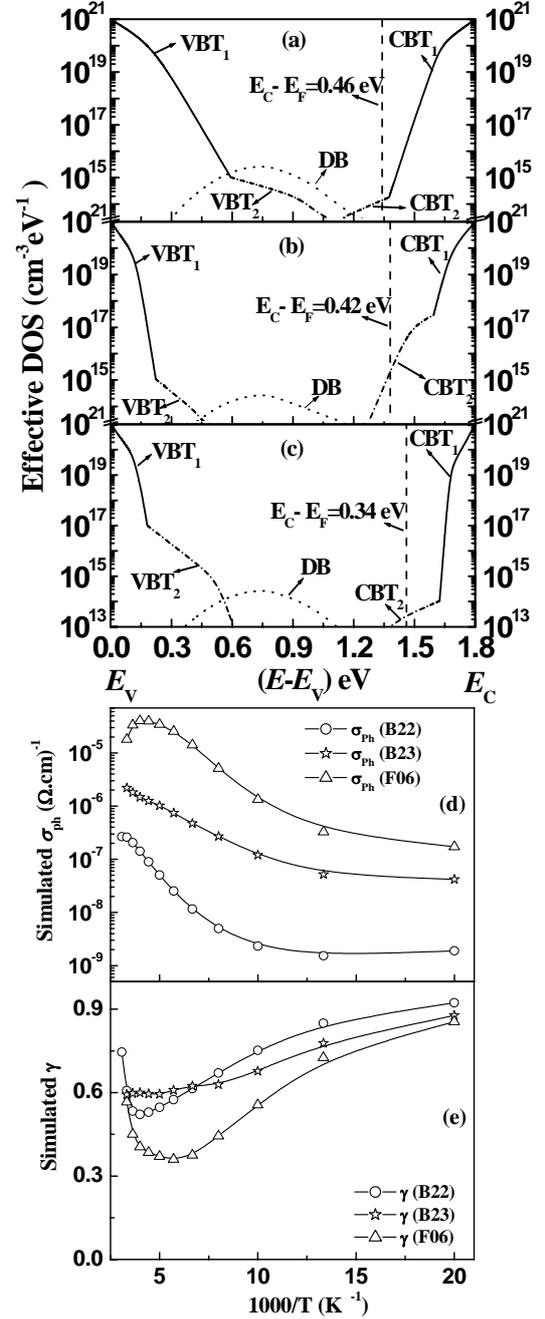

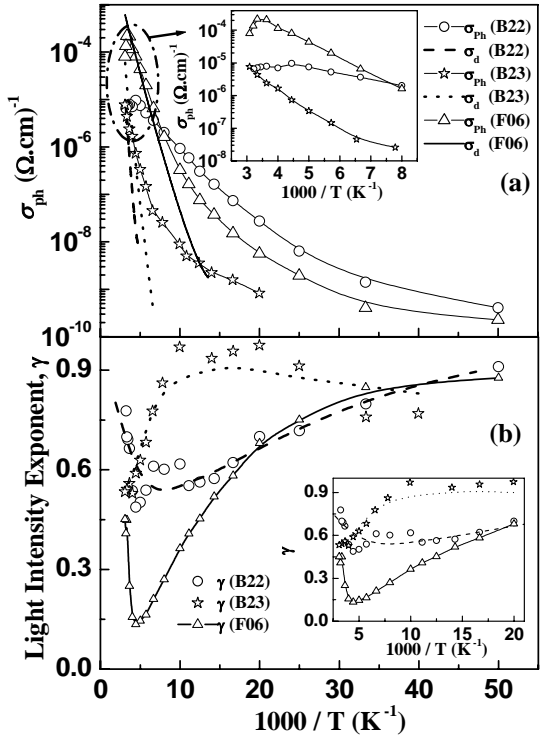

FIG. 1. (a) Temperature dependence of photoconductivity [$\sigma_{ph}(T)$] of sample #B22 (*type-A*), #B23 (*type-B*) and #F06 (*type-C*) for a particular light intensity ($\phi \approx 10^{16}$ photons/cm$^2$-sec $\approx$ $G_L = 10^{19}$ cm$^{-3}$s$^{-1}$) are shown by line+symbol, and $\sigma_d(T)$ of the same samples are shown in dashed, dotted and solid lines respectively. The inset shows the zoomed view of $\sigma_{ph}(T)$; (b) The temperature dependence of light intensity exponent $\gamma$ of the same samples. The inset shows the zoomed view of $\gamma(T)$.

FIG. 2. The effective DOS models corresponding to *types-A*, *B* and *C* $\mu$c-Si:H materials are shown in graphs (a), (b) and (c) respectively. Simulated phototransport properties of these three types of material using the DOS as shown in respective graphs calculated for light intensity $G_L = 10^{19}$ cm$^{-3}$s$^{-1}$ are shown in: (d) simulated temperature dependent $\sigma_{ph}(T)$ and (e) simulated temperature dependent $\gamma$.

In order to understand the experimentally observed phototransport behavior, and elicit information about the recombination processes, we carried out a numerical modeling of photoconductivity. For this, we first constructed the different possible effective DOS distributions of the $\mu$c-Si:H samples of the *types- A, B* and *C* (shown in Fig. 2(a), (b) and (c) respectively) by considering the qualitative analysis mentioned above, and then carried out rigorous numerical simulations with sensitivity analysis using Shockley-Read statistics in steady state conditions to determine the recombination process.[27] The basic formalism[3,12] and numerical methods are the same as we had used in our earlier reported work on numerical modeling,[5,6,7] but the parameter values and the DOS features have been further fine-tuned, generating simulated values much closer to the experimental data, and the recombination mechanisms deduced by modeling are corroborative with the qualitative analysis. We have not used any unphysical parameter values just for the sake of better quantitative agreement; the mobility values of electrons ($\mu_n$) used in this work were obtained from other experiments by us, and mobility values of holes ($\mu_p$) are reported values. We have not considered tunneling[28,29] in the simulation procedure, as carrier conduction is dominated by band tail transport in our material and tunneling process does not assume importance in our modeling study. Further, this process is important only at temperatures below 100K, while our interest lies in the temperature range above 200K, where the important phototransport processes like *TQ* and $\gamma(T)$ behavior are observed.

The simulated temperature dependence of $\sigma_{ph}$ and $\gamma$ of all the three types of samples are shown in Fig. 2(d) and 2(e) respectively. As evident from the graphs, the results of our model simulation of photoconductivity are in excellent agreement with the experimental findings. The simulated temperature dependences of recombination rates deduced from the model simulation is shown in Fig. 3, where we see that the dominant recombination takes place in shallow VBT states and is independent of temperature in *type-A* material. In contrast, in *type-C* material, recombination is mostly dominated by the shallower VBT states at lower temperatures, but the deeper VBT states start playing a role in the recombination process as the temperature increases. The trapped carrier densities in these deeper VBT states act as safe hole traps, which do not play a role in recombination at low temperatures, but govern the recombination process at higher temperatures. This elucidates well the cause of anomalous behavior experimentally observed in phototransport behavior of *type-C* material, and affirms that the underlying causes of *TQ* are different for *type-A* and *type-C* µc-Si:H material.

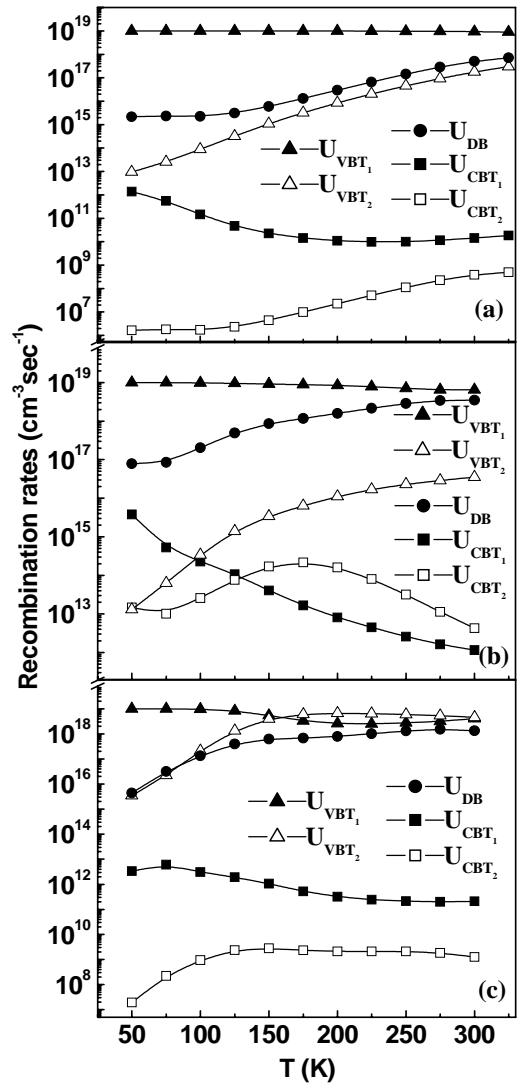

FIG. 3. Simulated recombination rates of (a) *type-A* $\mu$c-Si:H (b) *type-B* $\mu$c-Si:H and (c) *type-C* $\mu$c-Si:H. The generation rate is $G_L = 10^{19}$ cm$^{-3}$s$^{-1}$.

We have carried out simulation using a mobility gap ($E_g = E_c - E_v$) value of 1.8 eV, a value which is frequently used for a-Si:H, because, though the crystalline portion of the material affects the density of localized states in the gap, it is the defect states located in the *disordered phase* that actually take part in the recombination processes.[3] We had also carried out calculations with a mobility gap of 1.12 eV and DB distribution as suggested in the work by Lips *et al.*[30], which did not change the results of simulated phototransport behavior. This suggests that in highly crystalline µc-Si:H, the distribution of the localized states in the gap, rather than the gap width itself, influences the recombination processes, and thereby also the phototransport properties. That is, the defect states located within the disordered phase affect the recombination processes, wherever the disordered phase exists within the material (e.g., columnar or grain boundaries).

In summary, our study indicates that the morphological and microstructural differences in the different types of highly crystalline µc-Si:H material result in different phototransport behaviors. Our proposed effective DOS distributions successfully explain the phototransport properties in the light of microstructural properties, and are different for the three types of µc-Si:H materials, exhibiting structured band tails: a sharper, shallow tail originating from grain boundary defects and another less steeper deep tail associated with the defects in the columnar boundary regions, both of which have an exponential distribution. A band tail transport presents an experimentally and theoretically consistent picture of the transport in our µc-Si:H materials. The proposed DOS profile of *type-B* µc-Si:H is usable for any standard plasma deposited µc-Si:H, as it typifies the microstructure commonly possessed by such films.